\documentstyle[prl,epsf,aps,epsfig]{revtex}

\newcommand \be {\begin{equation}}
\newcommand \bea {\begin{eqnarray}}
\newcommand \ee {\end{equation}}
\newcommand \eea {\end{eqnarray}}
\newcommand \eps {\epsilon}
\newcommand \bi {\bibitem}
\newcommand \s {\sigma}

\def\Jp{J_{i_1 \ldots i_p}^{l_1 \ldots l_p}} 
\makeatletter
\newcommand\erfc{\mathop{\operator@font erfc}\nolimits}
\makeatother
\input{epsf}

\bigskip

\begin{document}

\twocolumn[\hsize\textwidth\columnwidth\hsize\csname@twocolumnfalse\endcsname
\draft       

\title{Complexity and line of critical points in a short-range spin-glass model}
\author{M. Campellone}
\address{ Dipartimento di Fisica and Sezione INFN,
Universit\`a di Roma ``La Sapienza''\\
Piazzale A. Moro 2, 00185 Rome (Italy)\\
e-mail:  matteo.campellone@roma1.infn.it}
\author{F. Ritort}
\address{Departament de Fisica Fonamental, Facultat de Fisica\\
Universitat de Barcelona, Diagonal 647\\
08028 Barcelona (Spain).\\
E-Mail: ritort@ffn.ub.es}

\date{\today}
\maketitle

\begin{abstract}
We investigate the critical behavior of a three-dimensional short-range
spin glass model in the presence of an external field $\eps$ conjugated
to the Edwards-Anderson order parameter. In the mean-field approximation
this model is described by the Adam-Gibbs-DiMarzio approach for the
glass transition.  By Monte Carlo numerical simulations we find
indications for the existence of a line of critical points in the plane
$(\eps,T)$ which separates two paramagnetic phases and terminates in a
critical endpoint. This line of critical points appears due to the large
degeneracy of metastable states present in the system (configurational
entropy) and is reminiscent of the first-order phase transition present
in the mean-field limit. We propose a scenario for the spin-glass
transition at $\eps=0$, driven by a spinodal point present above $T_c$,
which induces strong metastability through Griffiths singularities
effects and induces the absence of a two-step shape relaxation curve
characteristic of glasses.
\end{abstract} 

\vfill

\vfill

\twocolumn
\vskip.5pc
]

\narrowtext

Among all different approaches to understand the glass transition the
thermodynamic theory of Adam-Gibbs-Di Marzio (AGM) has deserved a lot of
interest during the last decades \cite{gima}. The AGM theory predicts
the occurrence, at a Kauzmann temperature $T_K$, of a second order phase
transition for the metastable undercooled liquid where the
configurational entropy (also called complexity) vanishes. The validity
of the AGM theory for real glasses has never been demonstrated so the
correct description of the glass transition still remains open
\cite{Angell}. An alternative dynamical approach was proposed in the
eighties to describe relaxational processes in the undercooled liquid
regime, experimentally observed in scattering and dielectric
measurements. This received the name of mode-coupling theory (MCT)
\cite{GO}.

Quite recently it has been realized that spin glasses are models which
account for both the thermodynamic (AGM) and the dynamical (MCT)
approaches \cite{KITHWO}. Although spin glasses are models with
quenched disorder (and structural glasses are not) this is not an
essential difference because the existence of a crystal phase in
structural glasses has no dramatic effect in the dynamics of the
(disordered) metastable undercooled liquid phase. Unfortunately, up to
now this connection between spin-glasses and glasses remains only at
the mean-field level and it is not clear what happens if one considers
short-range interactions. In fact, concepts such as complexity in AGM
or the ergodicity parameter in ideal MCT are originally mean-field and
it is not clear what is their relevance in short-ranged realistic
systems.


Recently it has been suggested \cite{CAFRPA,FRPA} that the effect of the
complexity could be observed through numerical simulations in a generic
glassy system coupling two replicas by introducing a term in the
Hamiltonian of the type $-\epsilon q$ ($\epsilon$ being the {\em
conjugate} field of the order parameter $q$ which is the overlap between
the configurations of the two replicas \cite{MPV}). Through the study of
an exactly solvable spin-glass model it has been shown the existence of
a first order transition line $T_c(\eps)$ with a critical end-point
\cite{FRPA}. Their result is an explicit check of the fact that the
glass transition for $\eps=0$ (where the complexity vanishes) is a first
order phase-transition (in the sense that the order parameter $q$ is
discontinuous) and the point $T_c=T_c(\eps=0)$ is a tricritical point.
Again, this result has been obtained within the mean-field approximation and
it is unclear to what extent this result is valid in a
finite-dimensional model. Recent numerical simulations on a short-range
version of $p$-spin Ising spin glass \cite{frpa2,mnpprz} have shown that
the mean-field discontinuous transition becomes continuous in finite
dimensions . So the characteristic first-order transition predicted
within mean-field theory dramatically changes in finite dimensions.  In
this work we want to show that this is not completely true and that some
features of the mean-field approximation indeed survive in finite
dimensions although they must be appropriately interpreted. This allows
to interpret our findings in terms of a picture for a continuous
spin-glass transition induced by the strong metastability and driven by
the collapse of the complexity or configurational entropy. Furthermore,
our results point in the direction that disordered systems in three
dimensions are very well described by a line of critical points
(characteristic of systems at their lower critical dimension) in
agreement with recent numerical simulations in the Edwards-Anderson
model in three dimensions \cite{CaPa}.

\begin{figure}
\begin{center}
\leavevmode
\epsfysize=230pt{\epsffile{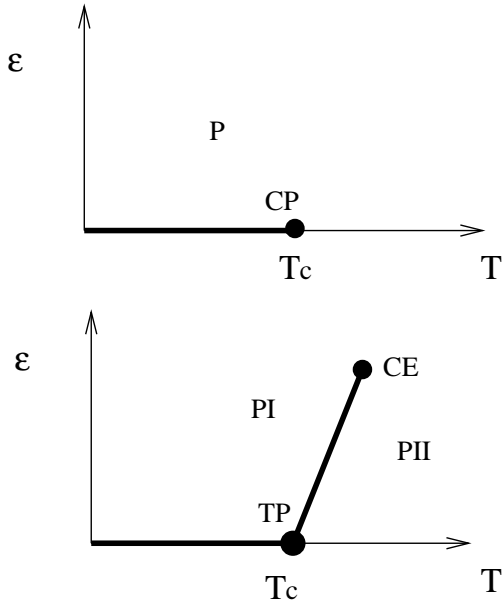}}
\end{center}
  \protect\caption[1]{ Different phases for the Edwards-Anderson model
(above) and the short-range p-spin model (below). In the former case
there is a single paramagnetic phase, in the latter two different
paramagnetic phases divided by a critical line which connects a
tricritical point (TCP) and a critical endpoint (CE) (for $p$ even the
line $TP-CE$ exists also for $\eps<0$). In both cases the spin-glass
phase is present only for $\eps=0$.  \protect\label{FIG1} }
\end{figure}

We shall consider a recently introduced short-range $p$-spin glass model
\cite{frpa2,mnpprz} which is defined on a $d$ dimensional hypercubic
lattice. On each site of the lattice there are $M$ spins interacting
with the following Hamiltonian

\be
{\cal H}_{p}(\{ \sigma \}) = -\sum_{<i_1,\cdots,i_p>}^{L^d}
 \sum_{l_1,\cdots,l_p=1}^M \Jp {\sigma_{i_1}}^{l_1}\cdots 
{\sigma_{i_p}}^{l_p}.\label{1}
\ee

By $\sum_{<i_1,\cdots,i_p>}^{L^{d}}$ we sum over all sites of the
lattice all possible groups of $p$ spins that can be formed between
spins on the same site and spins of adjacent sites.  In this work we
consider Ising spin variables. By ${\sigma_{i_r}}^{l_r}$ we denote the
$l_r^{th}$ spin of site $i_r$ with the index $l_{r}$ running from $1$
to $M$. For $p=2$ (\ref{1}) corresponds to the Edwards-Anderson
model. Although this model may seem quite artificial it has the
advantage of including multispin interactions in a lattice without
introducing new symmetries which may change the degeneracy of the
ground state \cite{AFR}.

Here we consider two identical coupled models (each defined through
(\ref{1})) via the following Hamiltonian

\be
{\cal H}_p^2(\{\s\},\{\tau\})={\cal H}_{p}(\{ \sigma \})+
{\cal H}_{p}(\{ \tau \})-\epsilon V q
\label{2}
\ee

where $V=L^3$ is the volume and $V q=\sum_{i=1}^V\s_i\tau_i$ defines the
order parameter. Because the $\eps=0$ transition in this model is
continuous we naively expect that (similarly as it happens in the
Edwards-Anderson model) there is no phase transition for $\eps\ne 0$. In
mean-field models with a one-step replica broken phase, a transition
exists for finite $\epsilon$ \cite{FRPA}. In that case, there is a
transition line which separates two paramagnetic phases with a finite
latent heat (which vanishes at the critical endpoint). The two possible
phases are depicted in figure~\ref{FIG1}. The Edwards-Anderson order
parameter $q_{EA}$ (defined as the infinite-time limit of the
equilibrium autocorrelation function \cite{MPV}) displays a finite jump
across the line $T_c(\epsilon)$ which vanishes at the critical
endpoint. Here we find strong indications, through Monte Carlo
simulations, that this first-order line $T_c(\eps)$ persists in finite
dimensions but becomes a line of {\em critical points}. So, in finite
dimensions the first-order line becomes continuous (i.e. $q_{EA}$ is
continuous when crossing the transition line and there is no latent
heat), the critical endpoint displaying a higher-order singularity.

\begin{figure}[t]
\begin{center}
\leavevmode \epsfig{file=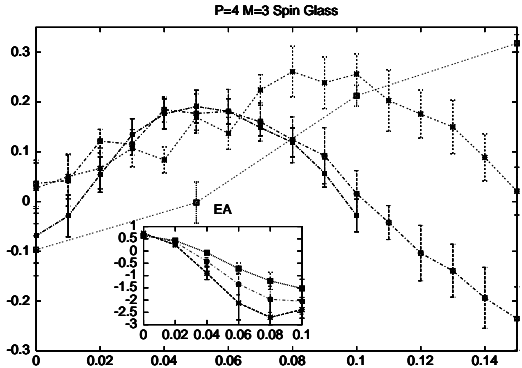,width=5truecm,angle=270}
\end{center}
\caption{Binder parameter as a function of $\eps$ at temperature
$T=3.2$ for different sizes ($L=3$ empty squares, $L=4$ filled
circles, $L=5$ empty circles, $L=6$ filled squares). The same plot for
the Edwards-Anderson model (inset) with $T=1.3,1.4,1.5$ ($T_c\simeq
1.2$ \protect\cite{YOUNG}) shows a completely different behavior
compared to the $p=4$ model.}
\label{FIG2}
\end{figure}

In this paper we will focus our research of (\ref{1}) for case $M=3, p=4$
in $D=3$. Measurements of the spin-glass susceptibility for (\ref{1})
show that this model has a continuous finite-temperature phase
transition at $T_c\simeq 2.6$ with a divergent spin-glass susceptibility
and a small negative specific heat exponent \cite{frpa2}. To evidenciate
a phase transition for finite $\epsilon$ which separates two different
paramagnetic phases we have done a detailed Monte Carlo study
\cite{DETAILS} of the Binder parameter as a function of both the
coupling $\eps$ and the temperature in the paramagnetic region. The
Binder parameter is usually defined through the relation,

\be
g(\eps,T)=\frac{1}{2}\bigl( 3-\frac{\overline{(q-<q>)^4}}
{(\overline{(q-<q>)^2})^2}\bigr)
\label{3}
\ee

where $<..>$ stands for statistical average and $\overline{(.)}$ for
disorder average. Concerning the Binder parameter we expect that it
should vanish everywhere in both paramagnetic phases except at the
critical line where it should be finite. So, if we fix $\eps$ (or,
equivalently the temperature) and vary the temperature (equivalently
$\epsilon$) we expect the presence a maximum approximately located on
the transition line at a temperature $T_c(L,\eps)$.  The results for
$g(\eps,T)$ are shown in figure~\ref{FIG2} for $T=3.2$ and different
sizes.  Note that the Binder parameter shows a maximum located at
$\eps=0.05$. As a comparison we have done simulations for the
Edwards-Anderson model in three dimensions above $T_c$ by coupling two
replicas which evidenciate the absence of a maximum in this case. So the
presence of a maximum in $g(\eps,T)$ already for small sizes is a main
feature of this model. For $L=3,4,5$ finite-size corrections appear to
be quite strong (this was already observed in \cite{frpa2} by measuring
the $P(q)$) and the position of the maximum of the Binder parameter as
well as its value both shift with $L$. Nevertheless, the maximum of the
Binder parameter for $L=6$ superimposes with the maximum for $L=5$ and
the Binder parameter goes to zero outside from the maximum as $L$
increases.  This result indicates the presence of a critical line which
separates two paramagnetic phases. A more stringent test of this result
requires simulating larger sizes than those we did. Unfortunately for
$L=7$ the thermalization time is much larger than what we can afford
with the present numerical methods. Actually, a test of the needed
thermalization time shows that it grows dramatically with $L$ and for
$L=7$ and $T=3.2$ this may be larger than several hundreds of million of
Msteps for a non vanishing fraction of disorder realizations. As we will
argue later this is consequence of the strong metastability and the
highly corrugated landscape characteristic of this model which, on the
other hand, induces the existence of this critical line (actually
thermalization is much easier in the Edwards-Anderson model).  Note that
the fact that the maximum value of $g$ along the critical line is
smaller than one (approximately 0.2) is an indication that the
transition is continuous in $q$ (if there were a finite jump in $q$ the
maximum value of $g$ should be 1). Moreover, our data do not show any
indication of a jump in the value $q(\eps)$ as a function of the
temperature (or $q(T)$ as a function of $\eps$) in the region of the
maximum of $g$.

\begin{figure}[t]
\begin{center}
\leavevmode \epsfig{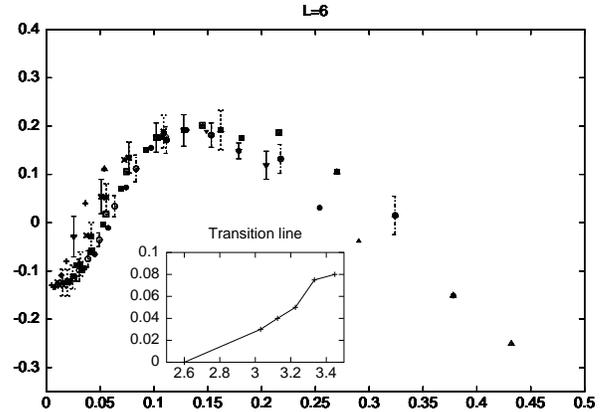}
\end{center}
\caption{Binder parameter plotted as a function of
$\eps(T-2.6)^{-2}$ in the range $.01\le \eps\le .09$, $3.0\le T\le 4.0$ for
$L=6$. In the inset we show the estimated critical line (see the text)
compared to the points for $L=5,6$ where appears a maximum of $g$ in the plane
$(\eps,T)$.}
\label{FIG3}
\end{figure}

In what follows we try to estimate the shape of this critical line using
finite-size scaling techniques. A detailed investigation of the critical
line as well as its critical exponents is presently out of reach due to
the smallness of the sizes studied. Still we can approximately determine
its shape. Let us suppose (as data of figure~\ref{FIG2} suggests) that
there is a critical line $\eps\sim C (T-T_c)^{\lambda}$ where
$C,\lambda>0$ and $T_c\simeq 2.60$. Assuming the validity of the scaling
hypothesis we may write $g(T,\eps)\equiv
\hat{g}(\eps(T-T_c)^{-\lambda})$.  In figure~\ref{FIG3} we plot the
scaling behavior within the scaling region for different values of $T$
and $\eps$ for the largest size $L=6$. The scaling is quite good and
proves two results: 1) The position of the maximum stays along a well
defined line in the $(\eps,T)$ plane and 2) The value of the maximum of
$g$ is the same everywhere along that line. A good collapse of data is
obtained with an exponent $\lambda\simeq 2$. The position of the maximum
of the scaled data yields a value of $C\simeq 0.17$ and $g_{max}\simeq
0.2$. This value is universal along the critical line and approximately
coincides with the value of the Binder parameter at the tricritical
point \cite{frpa2}. Although we expect that the critical line will have some $L$
dependence, very similar results obtained for $L=5$ indicate that our
estimate of the critical line for $L=6$ is a good estimate of the true
(infinite volume) line. 

\vspace{1.cm}

\begin{figure}
\begin{center}
\leavevmode
\epsfysize=160pt{\epsffile{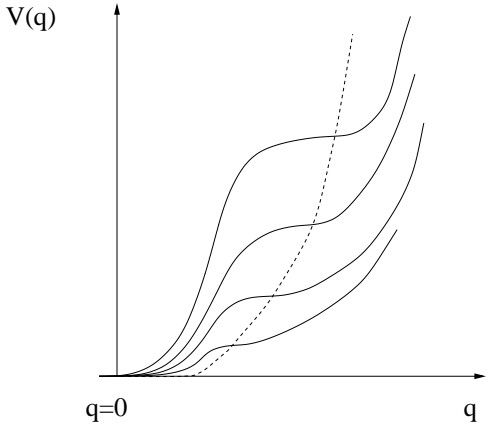}}
\end{center}
  \protect\caption[1]{ Potential function for the short-range p-spin
spin glass at $\eps=0$ above $T_c$ as a function of temperature (from
high to low temperatures from above to below). The secondary minimum at
$q\ne 0$ is a spinodal point. The height of that secondary saddle point
is the complexity (the logarithm of the number of metastable states per
site) which vanishes at $T_c$. The potential at $T=T_c$ becomes flat and
$q_{EA}$ vanishes at $T_c$ (following the dashed
line).\protect\label{FIG4} }
\end{figure}

Now we discuss the physical interpretation of this critical line.  As
already said in the introduction, the presence of this transition line
is consequence of the existence of an exponentially large number (with
the volume $\exp (\alpha V)$, $\alpha$ being a coefficient) of
metastable states in this model. In mean-field theory (AGM or ideal MCT)
metastable states have an infinite lifetime so there are infinitely
large barriers which separate them. This is the reason why in mean-field
theory ergodicity already breaks at the mode-coupling (also called
dynamical) temperature. In short-ranged systems or real glasses
metastable states decay by nucleation processes so ergodicity is always
restored. A typical feature of the ideal MCT singularity is the
characteristic two-step relaxational decay in correlation functions, the
so-called $\alpha$ and $\beta$ processes. In ideal MCT the typical
relaxation time associated to the $\alpha$ process diverges at $T_d$ and
the ergodicity parameter jumps discontinuously at $T_d$. An accurate
study of correlation functions reveals that the two-step characteristic
relaxation curve is absent in the present model (\ref{1})
\cite{frpa2,mnpprz}. The absence of a plateau in these curves indicate that do
not exist two well separated time-scales ($\alpha$ and $\beta$), like in
generic glass-forming liquids, but a continuous hierarchy of time scales
for nucleation processes. We interpret this result as consequence of the
continuous nature of the transition everywhere in the critical line
$\eps-T$.
A possible scenario for the potential function for this type of
transition is depicted in figure~\ref{FIG4}. There is not a typical
time scale for nucleation process (where a small excited droplet or
bubble decays from $q=q_{EA}$ to $q=0$) and the potential around the
secondary minimum $q=q_{EA}$ is marginally stable. Across the critical
line the Edwards-Anderson parameter $q_{EA}$ is continuous but
$\frac{dq_{EA}}{d\eps}$ is discontinuous, the potential being completely
flat in $q$. Note that at $T=T_c,\eps=0$ where the complexity vanishes,
the Edwards-Anderson parameter is also continuous in agreement
with the absence of the two-step relaxation in this model.
The complexity in this model is defined by the height of the secondary
saddle point which moves towards $q=0$ for $T=T_c$. The present scenario
is very different to that found in the Edwards-Anderson model in finite
dimensions where no additional spinodal point (apart from the
paramagnetic one $q=0$) is found above $T_c$.  Concerning dynamical
processes and thermalization effects this model behaves also quite
differently.  Spatial regions may be frozen if their local temperature
(measured always respect to the intensity of the local interaction) is
low enough for nucleation processes to decay very slowly. This is
apparent from figure~\ref{FIG4} where the secondary saddle point may
become a stable minima in certain Griffiths regions.  This effect has
been observed in numerical studies of the $P(q)$ where a secondary peak
in that distribution function has been observed already for small sizes
for certain disorder realizations \cite{frpa2}.

In summary, we have studied a short-range spin-glass model which in the
mean-field approximation is well consistent with the Adam-Gibbs-DiMarzio
theory and with the ideal mode-coupling theory. In the $\eps-T$ plane we
have found evidence for a line of critical points. The sole existence of
this transition suggests that a bunch of metastable states do change the
high-temperature behavior through their configurational entropy or
complexity.  The interest of this study is that the full analysis is
done in the high-temperature phase where thermalization is easier to
achieve (but still, huge thermalization times are needed for large
sizes) compared to the low-temperature region. To our knowledge, such an
analysis of this paramagnetic-paramagnetic transition has not been
previously done for spin glasses.  The main caveat of our results is
that we cannot exclude the possibility that we are observing a crossover
behavior where the maximum value of $g$ would eventually vanish for
$L\to\infty$ \cite{MBD}. Unfortunately, metastability slows very much
thermalization even in the high-temperature region.  Further studies
should extend the present analysis to the study of the potential energy
landscape and a numerical estimate of the configurational entropy for
this model (such as has been done for glass forming liquids \cite{SKT})
as well as the role of the Griffiths singularities in the
dynamics. Finally, because a finite $\eps$ breaks the $q\to -q$ symmetry
the existence of this critical line cannot be simply explained in terms
of a low-temperature spin-glass phase characterized by only two
symmetrically related equilibrium states. So probably one must invoke
more complex descriptions for the ground state structure and the type
of excitations in the spin-glass phase.



{\bf Acknowledgments}. We thank G. Parisi for discussions and
F. G. Padilla for a careful reading of the manuscript. This work has
been supported by the Ministerio de Educaci\'on y Ciencia in Spain
through the project PB97-0971.  M.C. thanks Fondazione Angelo Della
Riccia for financial support.

\hspace{-2cm}

\vfill

\end{document}